\documentclass[aps,twocolumn,preprintnumbers,showpacs,showkeys,nofootinbib%
]{revtex4}
\usepackage{epsfig}
\usepackage{amssymb,amsmath,amsfonts,amsthm,graphicx,psfrag}
\graphicspath{{./Figures/}}                 
\usepackage{array}
\usepackage{picture}
\usepackage{subfig}
\usepackage{graphicx}
\usepackage{textpos}

\newcommand{\beq}{\begin{equation}}
\newcommand{\eeq}{\end{equation}}
\newcommand{\bea}{\begin{eqnarray}}
\newcommand{\eea}{\end{eqnarray}}

\newcommand{\ds}{\displaystyle}

\newcommand{\aasymt}{{\cal A}}

\begin{document}

\title{The $\big\langle A^2 \big\rangle$ Asymmetry and Longitudinal Propagator
in Lattice $SU(2)$ Gluodynamics at $T\simeq T_c$.}

\author{V.~G.~Bornyakov}
\affiliation{
NRC ``Kurchatov Institute'' - IHEP, 142281 Protvino, Russia \\
School of Biomedicine, Far East Federal University, 690950 Vladivostok, Russia }

\author{V.~V.~Bryzgalov}
\affiliation{NRC ``Kurchatov Institute'' - IHEP, 142281 Protvino, Russia \\
}

\author{V.~K.~Mitrjushkin}
\affiliation{Joint Institute for Nuclear Research, 141980 Dubna, Russia \\
NRC ``Kurchatov Institute'' - ITEP, 117218 Moscow, Russia }

\author{R.~N.~Rogalyov}
\affiliation{ NRC ``Kurchatov Institute'' - IHEP, 142281 Protvino, Russia \\
}


\begin{abstract}
We study numerically the chromoelectric-chromomagnetic
asymmetry of the dimension two gluon condensate
as well as the longitudinal
gluon propagator at $T\simeq T_c$ in the Landau-gauge $SU(2)$ lattice
gauge theory.
We show that substantial correlation between the asymmetry
and the Polyakov loop as well as the correlation between the longitudinal
propagator and the Polyakov loop pave the way to studies
of the critical behavior of the asymmetry and the
longitudinal propagator. The respective values of
critical exponents and amplitudes are evaluated.
\end{abstract}



\maketitle

\section{Introduction}
\label{sec:introduction}

The gluon propagators were studied intensively in lattice gauge theories. 
Recently also the asymmetry of the chromoelectric-chromomagnetic
gluon condensate have received considerable attention.
Motivation for these studies can be found, e.g. in
\cite{Maas:2011se,Aouane:2011fv,Chernodub:2008kf,Vercauteren:2010rk} %
and references therein.
We only mention the relation of the zero-momentum
longitudinal and transverse propagators to the
chromoelectric and chromomagnetic screening masses
and, therefore, to the properties of strongly interacting quark-gluon matter.
Our attention here is concentrated on the critical behavior of these quantities
in the Landau-gauge $SU(2)$ lattice gauge theory.
It is well known that the second-order phase transition
of the 3D Ising universality class occurs in this model
and the Polyakov loop provides the order parameter \cite{McLerran:1981pb,Svetitsky:1982gs};
its nonzero value is associated with the spontaneous breaking of the
$Z_2$ center symmetry.

Though the behavior of the asymmetry and the gluon propagators
at $T\sim T_c$ have received much attention in the literature,
the situation with their temperature and volume dependence
in a close vicinity of $T_c$ is far from being clear.

It was shown in Ref.~\cite{Maas:2011ez} that the phase transition
both for $SU(2)$ and $SU(3)$ can be
clearly identified by a peak in the temperature derivative
of the chromoelectric screening mass
\beq
m_e={1\over \sqrt{D_L(0)}}
\eeq
where $D_L(0)$ is the zero-momentum longitudinal gluon propagator.
It was assumed that the critical behavior of $D_L(0)$
stems from the interference of the Gribov-copy effects
and the singular critical behavior
\beq
D_L(0)\sim |\tau |^{-\gamma_D}\;,
\eeq
where $\ds \tau={T-T_c\over T_c}$.
Some reasoning was given in favor of the
equality $\gamma_D =\gamma$, where $\gamma$ 
is the critical exponent characterizing the behavior of the
standard order-parameter correlator,
\beq
G(\vec 0)=\int d\vec x \langle {\cal P}(\vec x) {\cal P}(\vec 0)\rangle
\sim |\tau |^{-\gamma}\;,
\eeq
${\cal P}$ is the Polyakov loop.

Here we suggest a new approach to the studies of the
critical behavior of $D_L(0)$, which makes it possible to
overcome the difficulties associated with the need
to combine the Gribov mass with the singular critical behavior.
We argue that the critical exponent $\gamma$ is unrelated
to the critical behavior of $D_L(0)$.

Our approach is based on the study of correlations between
${\cal P}$ and $D_L(0)$ 
and between ${\cal P}$ and the asymmetry ${\cal A}$ which will be defined later in this section.
In the studies of these correlations
we employ well-established properties of the Polyakov loop.

In particular, the leading term of the asymptotic expansion
of the Polyakov loop in the infinite-volume limit
has the form
\beq
{\cal P} = B \tau^\beta  + \overline{o}(\tau^\beta)\; ,
\eeq
where the critical exponent $\beta$ coincides with that evaluated
in the 3D Ising model if the universality hypothesis is satisfied.
Recently, the critical exponents of the 3D Ising model were
computed with a great precision \cite{Kos:2016ysd},
\bea
\beta &=& 0.326419(3), \\ \nonumber
\nu &=& 0.629971(4), \\ \nonumber
\gamma&=& 1.237075(10). \nonumber
\eea
In the $SU(2)$ lattice gauge theory,
the critical amplitude $B = 0.825(1)$ was evaluated in \cite{Engels:1998nv}
in the case of $N_t=4$.

We study critical behavior of the quantities
\beq
{\cal Y} =  \aasymt - \aasymt_C
\eeq
and
\beq
{\cal D} = D_L(0) - D_L^C(0) \; ,
\eeq
where $D_L^C(0)$ is the average value of the longitudinal
gluon propagator at the critical temperature (referred to as $\ds{1\over m_{Gribov}^2} $
in \cite{Maas:2011ez}) and $\aasymt_C$ is the average value of the asymmetry at the
critical temperature in the infinite-volume limit.
Based on the analysis of correlations,
we argue that the leading term of the asymptotic expansion
of $\aasymt$ in $\tau$ at $\tau\to 0_+$
has the form
\beq
{\cal Y} \simeq C_{\cal Y} \tau^\beta\; ,
\eeq
where $\beta$ is the critical exponent of the $3D$ Ising model.
We also show that ${\cal D}$ has a similar behavior
\beq
{\cal D} \simeq C_{\cal D} \tau^\beta\; ,
\eeq
and evaluate the critical amplitudes $C_{\cal Y}$ and
$C_{\cal D}$.

\subsection{Definitions and simulation details}

At nonzero temperatures, there are electric
\beq
\big\langle A_E^2 \big\rangle = g^2 \big\langle  A^a_4 (x)  A^a_4 (x) \big\rangle, \quad
\eeq
and magnetic 
\beq
\big\langle A_M^2 \big\rangle = g^2 \big\langle  A^a_i (x)  A^a_i (x) \big\rangle.
\eeq
dimension two condensates.
The quantity of particular interest is the (color)electric-magnetic
asymmetry
\beq
 \aasymt \equiv {1\over T^2} \left(\big\langle A_E^2 \big\rangle
 - \frac{1}{3} \big\langle A_M^2 \big\rangle \right)\; ,
\eeq
it peaks at the phase transition and monotonically decreases
in the deconfinement phase \cite{Chernodub:2008kf}.
Expressions for $\aasymt$, definitions of the propagators $D_L(p)$ and $D_T(p)$,
and relations relations between them are given in \cite{Bornyakov:2010nc,Bornyakov:2016geh},
and \cite{Chernodub:2008kf}.

We study these quantities in the SU(2) lattice gauge theory with the standard Wilson action in the Landau gauge.

We generated ensembles of $O(1000)$ independent Monte Carlo
lattice gauge-field configurations on the $N_t\times N_s^3$ lattice
(in our study, $N_t=8$ and $N_s = 32\div 72$).
Consecutive configurations (considered as
independent) were separated by $100\div 450$
sweeps, each sweep consisting of one local heatbath update followed
by $N_s/2$ microcanonical updates.

Following Ref.~\cite{Bornyakov:2009ug}, we use the gauge-fixing algorithm
that combines $Z(2)$ flips for space directions with the simulated annealing
(SA) algorithm followed by overrelaxation. It is referred to as the `FSA' algorithm.
The other details of simulations and a more thorough description
of the gauge-fixing procedure can be found in \cite{Bornyakov:2016geh}.

Here we do not consider details of the approach to the continuum limit and renormalization
considering that the lattices with $N_t=8$ (corresponding to spacing $a\simeq 0.08$~fm)
are sufficiently fine. Thus we consider a bare gluon propagator and
the respective asymmetry.

\section{$A^2$ asymmetry in the deconfinement phase}

Details of the $\aasymt$ behavior near $T_c$ were omitted in the pioneering
work \cite{Chernodub:2008kf} because of low statistics.
Critical behavior of the asymmetry was first investigated in
\cite{Bornyakov:2015kwx}, however, that analysis
was based on configurations with positive Polyakov loop only.
Moreover, the distribution of the configurations in $\aasymt$
was not considered though the distributions in the Polyakov loop
were successfully employed in the studies of the
confinement-deconfinement phase transition.

Here we overcome the limitations of the
approach used in \cite{Bornyakov:2015kwx}.
Firstly, we compute the average value of the asymmetry
using simulation data in both $Z_2$ sectors and,
secondly, focus our attention on the distribution in $\aasymt$
and the correlation between the asymmetry $\aasymt$ and
the Polyakov loop ${\cal P}$.

To decide whether the average value of the asymmetry
depends on $\tau$, we fit it to a constant function,
\beq\label{eq:asym_const_fit}
\langle \aasymt \rangle \simeq c \;.
\eeq
The results are shown in Table~\ref{tab:asym_const_fit}.
In contrast to the conclusions made in \cite{Bornyakov:2015kwx},
the average value of the asymmetry
depends neither on the temperature nor on the lattice size.
\begin{table}[tbh]
\begin{center}
\vspace*{0.2cm}
\begin{tabular}{|c|c|c|} \hline
      &               &            \\[-2mm]
$N_s$   & ~~ $c$~~ &  ~$p$-value~  \\[-2mm]
      &               &             \\
   \hline\hline
32 & 4.67(4)   &  0.911 \\
48 & 4.61(3)   &  0.995 \\
72 & 4.60(2)   &  0.883 \\
\hline\hline
\end{tabular}
\end{center}
\caption{Results of the fit (\ref{eq:asym_const_fit}) over the range $-0.01< \tau < 0.03$.
}
\label{tab:asym_const_fit}
\end{table}

\begin{figure}[tbh]
\vspace*{-35mm}
\includegraphics[width=8cm]{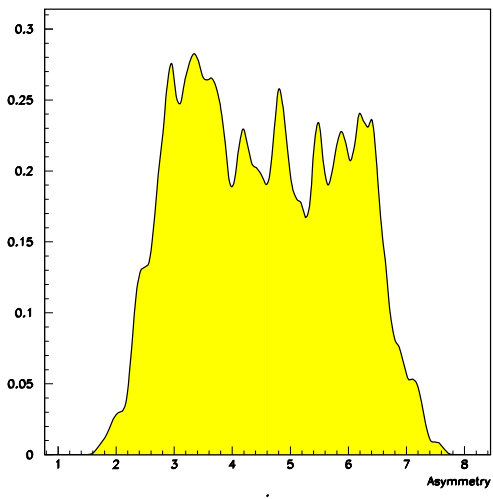}
\caption{Probability density function for the distribution in the asymmetry
at $\tau=-0.00765$, $L=6.0$~fm.}
\label{fig:asym_pdf_2508}
\end{figure}

\begin{figure}[tbh]
\vspace*{-35mm}
\includegraphics[width=8cm]{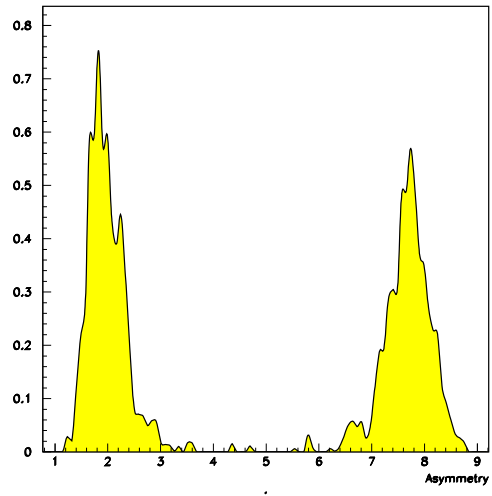}
\caption{Probability density function for the distribution in the asymmetry
at $\tau=0.02459$,  $L=5.8$~fm.}
\label{fig:asym_pdf_2518}
\end{figure}

In order to investigate the temperature dependence of the asymmetry
in more detail, we consider also the distribution of
generated configurations in the asymmetry.
First we notice that the distribution for $T<T_c$
(Fig.\ref{fig:asym_pdf_2508}) differs significantly
from that for $T>T_c$ (Fig.\ref{fig:asym_pdf_2518});
at $T>T_c$ two peaks emerge similar to the case of the Polyakov loop.

\begin{figure}[tbh]
\vspace*{-19mm}
\includegraphics[width=8cm]{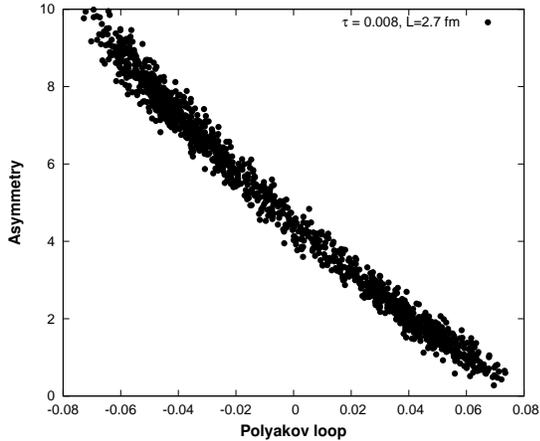}
\vspace*{-23mm}
\caption{Scatter plot demonstrating correlation between the asymmetry ${\cal A}$ and the Polyakov loop ${\cal
P}$; $\tau=0.008$; lattice size $L=2.7$~fm.}
\label{fig:asym_ploop_correl_2513_32}
\end{figure}

For this reason, we consider correlation between the asymmetry
and the Polyakov loop; this correlation is clearly seen, for example,
in the scatter plot in Fig.\ref{fig:asym_ploop_correl_2513_32}.
In view of such a substantial
correlation it is natural to consider
the asymmetry as the function of the Polyakov loop
and employ the regression analysis.

The regression analysis deals with the conditional distribution
described by the cumulative distribution function $F({\cal A}|{\cal P})$.
$F({\cal A}|{\cal P})$ describes the distribution of generated
configurations in the asymmetry for a fixed value ${\cal P}$ of the Polyakov loop.
We are interested in the conditional expectation
\beq
\langle {\cal A}\rangle_{\cal P}=E({\cal A}|{\cal P})=
\int {dF({\cal A}|{\cal P})\over d{\cal A}} \; {\cal A} d{\cal A}.
\eeq
In the neighborhood of the critical temperature
(that is, in the neighborhood of ${\cal P}=0$)
it can be fitted to a polynomial as follows
\beq\label{eq:asym_regressed_fit}
E({\cal A}|{\cal P}) \simeq {\cal A}_C + \sum_{j=1}^n A_j {\cal P}^j \;
\eeq

\begin{figure}[thh]
\vspace*{-19mm}
\includegraphics[width=8cm]{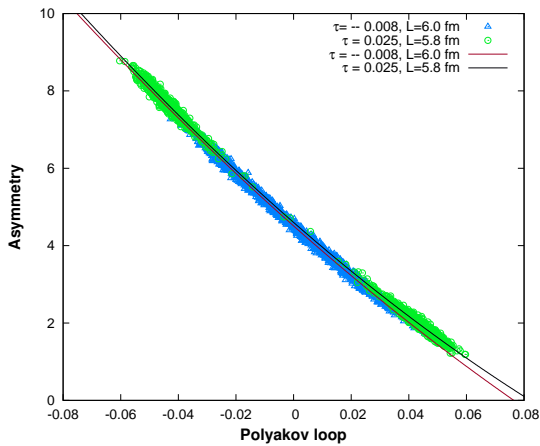}
\vspace*{-23mm}
\caption{Two scatter plots as in Fig.\ref{fig:asym_ploop_correl_2513_32}
on lattices $32^3\times 8$ at different values of $\tau$;
regression curves for both are shown; difference between them is within statistical errors.}
\label{fig:asym_ploop_corr_regr}
\end{figure}

For fixed value of $N_s$ we combined the data obtained for different values of $\ds {4\over g^2}\in [2.508, 2.518]$
($g$ is the coupling)
and determined the coefficients
$A_j$ using the method of least squares for the regression model with $n=3$.
The results are presented in Table~\ref{tab:asym_regressed_fit},
see also Fig.\ref{fig:asym_ploop_corr_regr}.

\begin{table}[tbh]
\begin{center}
\vspace*{0.2cm}
\begin{tabular}{|c|c|c|c|c|} \hline
      &            &             &            &  \\[-2mm]
$N_s$ & ~~ $A_0$~~ & ~~ $A_1$~~  & ~~ $A_2$~~ &  ~~ $A_3$~~ \\[-2mm]
      &            &             &            &  \\
   \hline\hline
32 & 4.4380(39) & -66.19(14)  &  141(2)  & 364(50) \\
48 & 4.4950(16) & -65.96(7)  &  136(1)  & 380(33) \\
72 & 4.5145(12) & -65.80(6)  &  139(1)  & 400(37) \\
\hline\hline
\end{tabular}
\end{center}
\caption{Results of the fit (\ref{eq:asym_regressed_fit})
over the range $-0.07 < {\cal P} < 0.07$.
The values $2.508\leq 4/g^2 \leq 2.518$ are taken into account.
}
\label{tab:asym_regressed_fit}
\end{table}

Now it is well to employ our knowledge of the
critical behavior of the Polyakov loop for the
investigation of the critical behavior of the asymmetry.
First we note that the width of the distribution
of field configurations in the Polyakov loop
tends to zero as the lattice size tends to infinity:
${\cal P}=\langle {\cal P}\rangle $\;. Thus the expectation value
$$
E\big({\cal A}\;\big|\;{\cal P}\!=\! {\cal P}(\tau)\big)
$$
determines the asymmetry  $\langle{\cal A}(\tau)\rangle$
in the infinite-volume limit.


Since at $\tau \leq 0$  \  \  $ {\cal P}(\tau) =0$,
the asymmetry furnishes a smooth function;
in some small interval $-\epsilon<\tau<0$
the asymmetry can be approximated by a constant,
\beq
{\cal A}\simeq {\cal A}_C\; ,
\eeq
in contrast to findings of \cite{Bornyakov:2015kwx}.

At $\tau>0$ spontaneous breaking of the
center symmetry occurs and we choose the positive
Polyakov-loop sector. In this sector, in the infinite-volume limit,
\beq
{\cal P}(\tau) = B\tau^\beta +{\overline o}(\tau^\beta)\; .
\eeq
Combining this relation with formula (\ref{eq:asym_regressed_fit}),
we arrive at the leading term of the asymptotic expansion
of the asymmetry at $\tau\to 0_+$,
\beq
\langle {\cal A}(\tau)\rangle = \; {\cal A}_C + A_1 B\tau^\beta +{\overline o}(\tau^\beta)\; ;
\eeq
therefore,
\beq\label{eq:CY_eq_BA1}
C_{\cal Y}=A_1 B.
\eeq
It should be emphasized that the only assumption
used in the derivation of this formula is that the relation
\beq\label{eq:osmall_hypothesis}
E({\cal A}|{\cal P}) = {\cal A}_C + A_1 {\cal P} + {\overline o}({\cal P})\;
\eeq
is valid at ${\cal P}>0$ with some values of ${\cal A}_C$ and $A_1$.

The critical amplitude $B$ in the case under consideration ($N_t=8$) can be
determined following the ideas outlined in \cite{Engels:1998nv}
and making use of the critical coupling $\ds {4\over g^2}=2.5104(2)$
reported in \cite{Velytsky:2007gj}.
With our statistics only a rough estimate can be obtained,
\beq\label{eq:B_estim}
B=0.147(35) \;,
\eeq
giving $C_{\cal Y} = -9.6(2.3)$, the error is almost completely accounted for
by poor precision in the determination of $B$, $A_1=-65.5(2)$
is determined much more precisely (here we discard errors associated
with the choice of the regression model etc).

\section{Critical behavior of $D_L(0)$}

\begin{figure}[tbh]
\vspace*{-19mm}
\includegraphics[width=8cm]{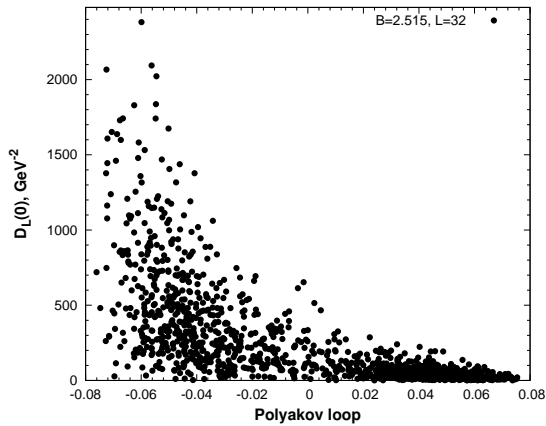}
\vspace*{-28mm}
\caption{Scatter plot demonstrating correlation
between $D_L(0)$ and the Polyakov loop ${\cal P}$;
$\tau=0.015$; lattice size $L=2.6$~fm.}
\label{fig:DL_ploop_corr_2515_32}
\end{figure}

We begin with the observation that the zero-momentum
longitudinal propagator (and, therefore, the electric screening mass
$\ds m_e={1\over \sqrt{D_L(0)}}$)
is strongly correlated with the Polyakov loop,
see the scatter plot in Fig.\ref{fig:DL_ploop_corr_2515_32}.
Thus we will study the dependence of $\mathbf{D} \equiv D_L(0)$ on the
Polyakov loop ${\cal P}$ near the criticality
on the basis of the regression analysis.

Our attention should be concentrated on the
cumulative distribution function
$F(\mathbf{D} |{\cal P})$
of the gluon propagator at a given value of the Polyakov loop ${\cal P}$
First we note that, at small $|\tau|$, $F(\mathbf{D}|{\cal P})$ does not depend
on $\tau$ and shows an approximate scaling with ${\cal P}$:
\beq
F(\mathbf{D}|s{\cal P})=F(\lambda(s)\mathbf{D}|{\cal P}),
\eeq
it is interesting that the distribution under consideration is non-Gaussian.

Similar to the case of asymmetry, we fit the combined data set
for all temperatures over the range $0.986T_c<T<1.03T_c$
to a common regression function
\beq\label{eq:D_L_fit_function}
\
\langle \mathbf{D} \rangle \simeq f({\cal P}) = \mathbf{D}_C
+ \sum_{j=1}^4 D_j {\cal P}^j \;,
\eeq
the zero-momentum propagator $\mathbf{D}$ is a predicted variable (regressand),
the Polyakov loop ${\cal P}$ is an explanatory variable (regressor).

\begin{figure}[tbh]
\vspace*{-19mm}
\includegraphics[width=8cm]{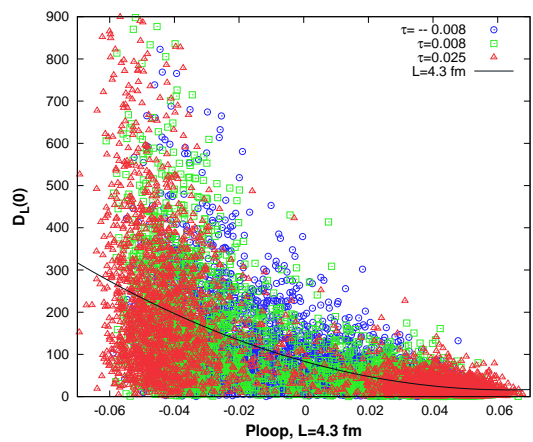}
\vspace*{-29mm}
\caption{Scatter plots as in Fig.\ref{fig:DL_ploop_corr_2515_32}
on lattices $48^3\times 8$ at different values of $\tau$;
common regression curve is shown.}
\label{fig:DL_ploo_corr_regressed}
\end{figure}

Thus in the infinite-volume limit
the leading terms of the asymptotic expansion at small $\tau$ have the form
\beq
\langle \mathbf{D} \rangle \simeq \mathbf{D}_C + \theta(\tau) C_{\cal D}  \tau^\beta
\eeq
where
\beq\label{eq:CD_eq_BD1}
C_{\cal D} = B D_1 = -\,330(80)\;\mbox{GeV}^{-2}\;,
\eeq
$B$ and $D_1$ appear in formulas (\ref{eq:B_estim}) and (\ref{eq:D_L_fit_function}), respectively;
again $D_1=-2200(100)$ is determined much more precisely than the critical amplitude $B$.

This result stems only from
\begin{itemize}
 \item the fact that, in the infinite-volume
limit, the distribution in ${\cal P}$ becomes infinitely narrow
(the Polyakov-loop susceptibility tends to infinity as $V\to\infty$);
\item and the smooth dependence of $\langle D_L(0) \rangle$ on ${\cal P}$ at ${\cal P}=0$.
\end{itemize}

In a series of studies \cite{Cucchieri:2012nx,Mendes:2014gva,Mendes:2015jea}
it was concluded that, in a large, however finite, volume,
$D_L(0)$ has a smooth behavior at the criticality:
it approaches a maximum value
at some temperature $T_m<T_c$ and then gradually decreases.
Such behavior is seen not only at $N_t=16$, as was stated in the
mentioned studies, but also at $N_t=8$ \cite{Bornyakov:2015kwx}.
This observation should have a natural explanation in our approach
provided that we consider only the positive Polyakov-loop sector
(which is the case for the cited works).

Firstly, we note that at the temperatures well below $T_c$
the longitudinal gluon propagator slowly increases with temperature.
We consider that this growth is unrelated to the observed correlation
with the Polyakov loop and we disregard reasons behind it.
Secondly, the values of the Polyakov loop are concentrated
near zero; however, when the temperature comes to the critical one from below,
the width of the distribution in the Polyakov loop
begins to rise. In view of the correlation shown in
Figs.\ref{fig:DL_ploop_corr_2515_32}~and~\ref{fig:DL_ploo_corr_regressed},
this results in a decrease of $D_L(0)$
in the positive Polyakov-loop sector (in the negative Polyakov-loop sector
$D_L(0)$ increases).
At $T=T_m$ the decrease caused by fluctuations near criticality
overwhelms the above-mentioned increase characteristic for the confinement domain.

It is not surprising that $D_L(0)$ shows a smooth behavior even at very large volumes
\cite{Cucchieri:2012nx,Mendes:2014gva,Mendes:2015jea}
because an approach to the infinite-volume limit
near the criticality presents a challenge, especially in the case when
only the positive Polyakov-loop sector is taken into consideration
\cite{Bornyakov:2015kwx}.
In any case, for a comprehensive investigation
of the critical behavior of Green functions
both Polyakov-loop sectors should be taken into account
in some neighborhood of the critical temperature.

\section{Conclusions}

We have studied the asymmetry $\aasymt$ and the
longitudinal gluon propagator
in the Landau-gauge $SU(2)$ gluodynamics on lattices with $32\leq N_s\leq 72$
and $N_t=8$ in the range of temperatures $0.99 T_c < T < 1.03 T_c$.
Our findings can be summarized as follows:

\begin{itemize}
\item Both the asymmetry and the longitudinal propagator have a
significant correlation with the Polyakov loop.
\item Regression analysis reveals that
$\langle {\cal A} \rangle_{\cal P}$ and $\langle D_L(0) \rangle_{\cal P}$
are smooth functions of the Polyakov loop. Therefore, if
the relation (\ref{eq:osmall_hypothesis}) is valid, the critical exponents
of $\aasymt - \aasymt_C$ and $D_L(0) - D_L^C(0)$ are given by
\beq\label{eq:main_result}
\beta_A = \beta_D = \beta = 0.326419(3)
\eeq
and the respective critical amplitudes are
given by the formulas (\ref{eq:CY_eq_BA1}) and (\ref{eq:CD_eq_BD1}).
Since the regression analysis gives only an indication
to the validity of (\ref{eq:osmall_hypothesis}),
more thorough investigation of smoothness of
$\langle {\cal A} \rangle_{\cal P}$ and $\langle D_L(0) \rangle_{\cal P}$
is needed.
\item Volume dependence of both $D_L(0)$ and ${\cal A}$
can be accounted for by the volume dependence of the Polyakov loop.
\end{itemize}
Similar studies of correlations between the Ployakov loop
and the gluon propagators in QCD may shed light on both critical behavior
of screening masses and dynamics of gluon fields near $T_c$.

\acknowledgments{Compuations were performed using IHEP (Protvino)
Central Linux Cluster and ITEP (Moscow) Linux Cluster.
The work was supported by the Russian Foundation for Basic Research, grant no.16-02-01146~A.}


\end{document}